\documentclass[twocolumn,amsmath,amssymb,aps,pra,floatfix]{revtex4-2}

\usepackage{physics}
\usepackage{amsmath}
\usepackage{cases}
\usepackage{url}
\usepackage{textcase}
\usepackage{amssymb}
\usepackage{graphicx}
\usepackage{bm} 

\usepackage{times}
\usepackage{float}
\usepackage{multirow,microtype,color,relsize}
\usepackage{subfigure}
\usepackage[breaklinks=true]{hyperref}
\usepackage[utf8]{inputenc}
\usepackage[english]{babel}
\hypersetup{colorlinks=true,linkcolor=blue,citecolor=blue}
\hypersetup{linktocpage}
\usepackage{CJKutf8}
\usepackage{breakcites}
\usepackage{float}
\usepackage{siunitx}
\usepackage[dvipsnames]{xcolor}
\definecolor{mypine}{RGB}{1, 121, 111}

\begin{document}
\title{Half-century of Efros-Shklovskii Coulomb gap. Romance with Coulomb interaction and disorder.}
\author{B. I. Shklovskii}
\affiliation{School of Physics and Astronomy, University of Minnesota, Minneapolis, Minnesota 55455, USA}
\date{\today}

\begin{abstract}
The Efros-Shklovskii (ES) Coulomb gap in the one-electron density of localized states and the ES law of the variable range hopping conductivity were coined 50 years ago. The theory and its first confirmations were reviewed in the SE monograph published 40-years ago. This paper reviews the subsequent experimental evidence, theoretical advancements, and novel applications of the ES law. Out of hundreds of experimental validations of the ES law in a diverse range of materials, I focus on those where the dynamic range of conductivity exceeds four orders of magnitude. These include three- and two-dimensional semiconductors under both zero and high magnetic fields, localized phases in the quantum Hall effect, granular metals, nanocrystal arrays, and conducting polymers. Additionally, I discuss the non-ohmic ES law and the Coulomb gap near insulator-metal transition. Recent developments of other concepts of the SE book are also discussed.
\end{abstract}

\maketitle

\section{Introduction}

50 years ago Efros and Shklovskii (ES) penned the paper~\cite{ES1975} “Coulomb gap and low temperature conductivity of disordered systems”. It elucidated how Coulomb interaction between localized electrons in semiconductors creates the power law Coulomb gap of the one-electron density of states near the Fermi level and changes temperature dependence of Mott variable hopping conductivity to the ES law. 
 
Localized electron states in lightly doped semiconductors are formed by randomly positioned donors. Typically semiconductors are compensated, i. e. $n$-type semiconductor with concentration of donors $N_D$ has a concentration $N_A < N_D$ of acceptors. At low temperatures each acceptor captures an electron from a donor and becomes negatively charged. As a result $N_A$ donors become positively charged while $N_D-N_A$ donors are still neutral. Charged donors and acceptors create a random potential which shifts energy of the donor $i$ up or down by a value $\epsilon_i$. This broadens the originally $\delta$-functional peak of the density of donor states. At low temperature lower energy occupied donor levels are separated by the Fermi level (chemical potential) $\mu$ from higher empty levels. Before 1975 it was believed that near the Fermi level $\mu$ the density of states (DOS) $g(\epsilon) = g(\mu) = Const$. ES showed~\cite{ES1975} that due to the Coulomb interaction between electrons 
\begin{equation} 
g(\epsilon) = \frac{\pi}{3} \frac{\kappa^3(\epsilon-\mu)^2}{e^6}
\label{eq:DOS3}
\end{equation}
in a three-dimensional (3D) semiconductor and 
\begin{equation} 
g(\epsilon) = \frac{\pi}{2} \frac{\kappa^2|\epsilon-\mu|}{e^4}
\label{eq:DOS2}
\end{equation}
in a two-dimensional (2D) one. Here $\kappa$ is the dielectric constant of the semiconductor (and in 2D of its environment), $e$ is the proton charge and coefficients $\pi/3$ and $\pi/2$ were found with help of the self-consistent approximation~\cite{Efros1976}. Thus, the density of states has a power law gap at the Fermi level, which is termed the Coulomb gap by ES. One can say that the Coulomb gap originates due to the exciton effect, i. e. attraction between an electron excited from occupied to an empty in the ground state donor to the hole it left at the original donor. The major consequence of this gap is replacement of the Mott variable range hopping conductivity law
\begin{equation} 
\sigma = \sigma_0 \exp{-\left(\frac{T_M}{T}\right)^{p}}
\label{eq:Mott3}
\end{equation}
with $p=1/4$ for three-dimensional (3D) and $p=1/3$ for 2D cases
by the ES law
\begin{equation} 
\sigma = \sigma_0 \exp{-\left(\frac{T_{ES}}{T}\right)^{1/2}}
\label{eq:ESlaw}
\end{equation}
valid for both 3D and 2D cases. Here $T$ is the temperature in energy units, $T_M = 22/g(\mu) \xi^3$ for 3D and $T_M = 13/g(\mu) \xi^2$ for 2D, $\xi$ is the localization length of localized states at the Fermi level (for a lightly doped semiconductors $\xi$ is equal to the Bohr radius of a donor $a$), and
\begin{equation} 
T_{ES} = \frac{C e^2}{\kappa \xi}, 
\label{eq:TESl}
\end{equation}
where $C = 2.8$ in 3D and $C = 6.2$ in 2D~\cite{SEbook,Nguyen1984}. 

Derivations of Eqs. (\ref{eq:DOS3}) and (\ref{eq:ESlaw}) and first experimental and numerical conformations of them are discussed at length in Shklovskii-Efros (SE) book~\cite{SEbook}. The Russian original of SE book was published in 1979, while the updated and extended English translation was published in 1984, 40 years ago. SE book is still widely used. I devote this article to the double 50/40 years anniversary. The main part of the article reviews new observations, generalizations and applications of Eqs. (\ref{eq:DOS3}) and (\ref{eq:ESlaw}) while the second part briefly mentions developments of other ideas of SE book. I try to avoid repetitions of the SE book. 

The plan of this short article is as follows. I dwell on major experimental confirmations of the ES Coulomb gap and ES law in Section II. Section III is devoted to non-Ohmic generalization of the ES law. In Section IV I discuss new advances of the Coulomb gap and ES law theories. I focus on the evolution of the Coulomb gap near metal-insulator transition in Section V. Section VI goes beyond the Coulomb gap and SE book and dwells on related problems of Coulomb interaction and disorder physics. I comment on developments of ideas of specific chapters of the SE book in Section VII. 

\section{New experimental confirmations of Coulomb gap and ES law}

An experimental verification of ES law Eq. (\ref{eq:ESlaw}) is done by plotting $\ln\sigma$ as a function of $T^{-1/2}$ and checking whether this plot can be fitted by a straight line. In order to reliably discriminate the ES law from Eq. (\ref{eq:Mott3}) with $p=1/4$ and particularly in 2D case with $p=1/3$ the dynamic range of measured conductivity should cover 2-3 orders of magnitude. In this review I dwell only on the most spectacular data which cover at least 4 orders of magnitude. 

I start from observations of the ES law in 3D elementary semiconductor crystals of Ge and Si. First high quality data for compensated germanium were discussed Chapters 9 and 10 of the SE book. Later the ES law was observed in neutron-transmutation-doped $p$-Ge in 4 orders of magnitude dynamic range of resistance (4 OMDRR)~\cite{Itoh1993} 
and even in 6 OMDRR~\cite{Schoepe1988}. Similar high quality data were obtained in Si samples in 5 OMDRR~\cite{McCammon2005}. It is remarkable that the data~\cite{Itoh1993,Schoepe1988,McCammon2005} were obtained on Ge and Si based bolometers and termistors for astrophysics and dark matter research. 

Other spectacular data for the ES law were obtained in the mixed valence cuprate LiCu$_3$O$_3$ in 7 OMDRR~\cite{Moser7}, in CdSe nanocrystal arrays in 6 OMDRR~\cite{Sion6}, in ZnO nanocrystal films in 4 OMDRR~\cite{Ben4}, in magnetite nanocrystal arrays in 7 OMDRR~\cite{zheng7}, in  A$_2$MnReO$_6$ ceramics~\cite{Ceramics5} in 5 OMDRR,
in granular metals in insulator in 4-8 OMDRR~\cite{Abeles,Deutscher}, in different conducting polymers~\cite{Zuppiroli10,Epstein6,polymer6,Heeger4,polymer5,polymer4,Chrys4} in 4-10 OMDRR, in graphene oxide thin films~\cite{Saiful4,Rudenko7} in 4-5 OMDRR, in thick multilayer disordered carbon nanotubes~\cite{Feldman} in 5 OMDRR, and in arrays of self assembled Ge quantum dots embedded into Si MOSFET~\cite{Yakimov} in 5 OMDRR.

Due to scattering of a tunneling electron on intermediate donors the ES law works even in strong magnetic field $B$~\cite{Boris1982}. 
This happens because this scattering changes the asymptotic behavior of a donor electron wave function in the orthogonal to magnetic field plane from the standard $\exp(-\rho^2/4\lambda^2)$ to $\exp(-\rho/L\lambda)$, where $\lambda = (\hbar c/eB)^{1/2}$ is the magnetic length and $L$ is the logarithmic function of $B$. Indeed, the data obtained on InSb samples~\cite{Tokumoto} (see SE Fig. 9.2b) demonstrates excellent agreement with the ES law in 7 OMDRR. Similar magnetoresistance  data were obtained for Ge samples in 4 OMDRR~\cite{Schoepe1988}.

Let us now switch to 2D systems with localized electron states. The ES law was observed in 6 OMDRR in the gated GaAs/GaAlAs low mobility heterostructures in a strong orthogonal magnetic field~\cite{Arnie6} which shrinks localized state wave functions and makes electron hops shorter. At smaller magnetic fields, where the hop length becomes as large as the distance to the gate the authors observed the transition from ES law (\ref{eq:ESlaw}) to 2D Mott law (\ref{eq:Mott3}) with $p=1/3$. They interpreted this transition as a result of screening of electron-electron interactions in two-dimensional electron gas (2DEG) by the metallic gate. This observation confirms that the ES law results from the Coulomb interaction between electrons. Screening induced crossover from ES to Mott law was also observed in arrays of self assembled Ge quantum dots~\cite{Yakimov}.  

Discovery of the Quantum Hall effect (QHE) in higher mobility samples lead to observations of ES law in 4 OMDR of the diagonal conductivity $\sigma_{xx}$ at integer filling factors $\nu$, where the electron states at the Fermi level are localized~\cite{Klaus4,Briggs4}. Ref.~\cite{polyakov93} argued that at low enough temperatures the ES law should be observed not only at integer $\nu$, but everywhere between two neighboring half-integers $\nu_c$ where electron states are still localized. The localization length $\xi$ diverges at a Landau level as $\xi \propto |\nu-\nu_c|^{-\gamma}$ with $\gamma=7/3$ leading to vanishing $T_{ES}(\nu)$ at all half-integers. At a finite $T$ equation $T_{ES}(|\nu-\nu_c|) = T$ defines the temperature dependent half-width $|\nu-\nu_c|\propto T^{1/\gamma}$ of a the peak of the resistivity $\rho_{xx}$ near $\nu_c$ in agreement with experimental data~\cite{Wei1988,Wei1992,Koch1992}.

Predictions of Ref.~\onlinecite{polyakov93} were also confirmed by comprehensive studies of ES variable range hopping in cleaner samples of GaAs~\cite{Furlan,Haug5} and graphene~\cite{Glattli}. In these papers ES law for $\sigma_{xx}$ was observed in 4-5 OMDRR and the localization length $\xi(\nu)$ was extracted from measured values of $T_{ES}$. It was found to diverge at $\nu_c$ with the theoretical critical index 7/3. Ref.~\onlinecite{Glattli} also studied transition from ES to Mott law due to screening of Coulomb interaction by the gate, again confirming the Coulomb origin of ES law. 

In the absence of screening the ES law may yield to the Mott law only when the bare (non-Coulomb) DOS $g(\mu)$ is very small, so that the Coulomb gap width determined by equating Eqs. (\ref{eq:DOS3}) and (\ref{eq:DOS2}) to $g(\mu)$ is narrower than the band of energies around the Fermi level involved in the variable range hopping. Such a situation takes place in amorphous semiconductors, where the Mott law was originally observed. Even there the ES law should work at low enough temperatures, but this requires measuring unrealistically large resistances. Approximate shapes of the crossover between Mott to ES laws where suggested in Refs.~\cite{Lien,Amnon}.

So far we talked about the integer QHE. The ES law was observed also in middles of the fractional QHE plateaus~\cite{Boebinger}. One can interpret such an observation assuming that fractional charges $q$ (say $q = -e/3$) are bound to donors and have the Coulomb gap of DOS. The variable range hopping of these fractional charges then obeys the ES law with $T_{ES} = Cq^2/\kappa\xi$, where $\xi$ is the localization length of of the fractional charge $q$. Small values of $q$ lead to a very small $T_{ES}$ in agreement with experiment~\cite{Boebinger}. Then, a straightforward extension of the logic of Ref.~\onlinecite{polyakov93} leads to the ES law for $\sigma_{xx}$ everywhere in a fractional plateau. The divergence of the localization length at a border between two neighboring fractions leads to the prediction of the same behavior of widths of their $\rho_{xx}$ peaks with growing temperature as in the integer QHE. This prediction agrees with experimental results~\cite{Engel,Muraki,Shayegan,Bid23} .

\section{Non-Ohmic ES law}

So far we discussed the Ohmic variable range hopping conductivity in small electric fields. It was shown theoretically~\cite{Shklov1973} that at very low temperatures the strong electric field $E$ acts as the effective temperature $T_E = \gamma e\xi E$, where $\gamma = 1/2$. The origin of $T_E$ can be understood using the concept of the electron quasi-Fermi level $\mu(x) = \mu(0) - eEx$ created by the electric field $E$ directed oppositely to the $x$-axis. Then an electron hopping from an occupied donors at $x=0$ with the energy close to $\mu(0)$ to an empty donor at the distance $x$ finds itself above the local Fermi level $\mu(x) = \mu(0) - eEx$ by energy $\epsilon = eE|x|$. Probability of such a hop is $P = \exp(-2x/\xi) = \exp(-\epsilon/T_E)$ determines the generation rate of electrons with energy $\epsilon$. Assuming that their energy relaxation rate is a weaker function of energy we arrive at the Boltzmann energy distribution with $T_E = e\xi E/2$.  

As a result Eqs. (\ref{eq:DOS3}) and (\ref{eq:DOS2}) lead to the exponential growth of the current $I$ with $E$ (the non-Ohmic ES law) 
\begin{equation} 
I = I_0 \exp{-\left(\frac{E_0}{E}\right)^{1/2}},
\label{eq:NOESlaw}
\end{equation}
where 
\begin{equation} 
E_0=\frac{2T_{ES}}{e\xi}= \frac{2Ce}{\kappa \xi^2}. 
\label{eq:E0}
\end{equation}
Fitting the lowest temperature data with Eq. (\ref{eq:NOESlaw}) and the temperature dependence of the Ohmic conductivity with Eq. (\ref{eq:ESlaw}) and using the resulting $T_{ES}$ and $E_0$ one can find unknown $\xi$ and $\kappa$. This is particularly important~\cite{polyakov93} in 3D cases where most measurements are done close to the insulator-metal transition at which both $\xi$ and $\kappa$ diverge~\cite{SEbook}. The non-Ohmic ES law was observed in large dynamic ranges of currents in Refs.~\onlinecite{Sion6,Saiful4,Rudenko7}. These authors then found both  $\xi$ and $\kappa$. 

Numerical simulation of 3D systems~\cite{Marianer1992,Nenashev11} arrived at the effective temperature $T_E = \gamma e\xi E$ both for electron energy distribution and for hopping conductivity with coefficients  $\gamma$ varying from 0.5 to 0.7. I do not know numerical results for 2D case, but I believe that in this case $\gamma \simeq 0.5$ as well. In the peculiar 1D case the value $\gamma = 1/8$ was analytically obtained in Ref.~\onlinecite{Fog}.

Eq. (\ref{eq:E0}) was also generalized for the low temperature variable range hopping in the quantum Hall effect, where the role of electric field $E$ is played by the Hall electric field~\cite{polyakov93}. This allowed to predict that the half-width $|\nu-\nu_c|$ of $\rho_{xx}$ peaks at a half-integer $\nu_c$ grows as $I^{1/2\gamma}$ with the current $I$. This prediction was confirmed by Refs.~\onlinecite{Furlan,Haug5,Glattli,Wei1994}. Temperature and current dependencies of $\rho_{xx}$ peak widths based on the ES law~\cite{polyakov93} provide the Coulomb microscopic interpretation for the quantum transition point scaling theory~\cite{Sondhi97}.

\section{Coulomb gap and ES law: new theoretical developments} 

Multiple numerical studies of the Coulomb gap density of states confirmed power laws Eqs. (\ref{eq:DOS3}), (\ref{eq:DOS2}) 
for three- and two-dimensional systems~\cite{Davies,Mobius92,Mobius2009,Mobius2010,Goethe,Palassini}. However, they did not find exponential depletion of the one-electron density of states at smallest energies due to the electron polaron effects predicted for three-dimensional case~\cite{Efros1976} and discussed in Chapter 10 of the SE book. This puzzle was resolved in Ref.~\onlinecite{Brian2011}. We argued that polaron effects are delayed till very small $|\epsilon-\mu|$ which can not be modelled with modern computers. In two dimensions the ES law was confirmed by the direct numerical modelling of the variable range hopping conductivity~\cite{Tsygankov}. 

Coulomb gap at finite temperatures is numerically studied in Refs.~\onlinecite{LevinT,Vojta}. It is found in Ref.~\onlinecite{LevinT} that the density of states at the Fermi level $g(\mu, T) \propto T^2$ in 3D and $g(\mu, T) \propto T$ in 2D in agreement with theoretical predictions, while in Ref.~\onlinecite{Vojta} both powers are found to be somewhat larger. 

Coulomb gap and variable range hopping are different~\cite{Logarithm} in a film with a very large dielectric constant in small dielectric constant environment where Coulomb potential logarithmically depends on distance. Because $\ln x$ can be replaced by $x^{1/3}$ with accuracy 15\% in the range $3 < x < 100$  standard ES arguments lead to $g(\epsilon) \propto (\epsilon-\mu)^{5}$ and $\ln(\sigma/\sigma_0) \propto T^{-3/4}$ in this case.

As we saw in Sec. II the ES law was observed in many granular metals and nanocrystal arrays~\cite{Abeles,Deutscher,Sion6,Ben4,Lopatin}. 
The theory of the Coulomb gap and the ES law for such systems was developed later than for semiconductors because of missing understanding of the origin of disorder. We argued that in the ground state of an array of metal granules immersed in an insulator, random charges of granules are made by addition of fractional image charges induced by randomly positioned neighboring charged impurities in the insulator~\cite{Zhang2004,Tianran2012}. 
The Coulomb gap physics of an immersed in an insulator array of superconducting granules was studied in Refs.~\onlinecite{Gal,Tianran20121}. Here one can observe the interplay of two Coulomb gaps due to repulsion of electrons and Cooper pairs. 
The role of charged impurities in a system of isolated long metallic needles or flexible wires dispersed and frozen in an insulator was studied in Ref.~\onlinecite{Tao2006,Tao20061}.

On the other hand, in periodic epitaxially connected semiconductor nanocrystal arrays disorder originates from random numbers of donors and acceptors in individual nanocrystals~\cite{Brian2012,Han2016,Uwe2016}. 
An interesting, but more complicated case of quasi-one-dimensional crystals with impurities was theoretically considered in Ref.~\onlinecite{Fogler2004} and experimentally studied in Ref.~\onlinecite{Babic}.

As we saw above, the ES law data allows us to find the localization length of localized states with energy close to the Fermi level. This required the development of a new theory of the localization length. It was shown that in granular metals the long distance tunneling happens through a chain of neighboring granules or, in other words, via relay co-tunneling~\cite{Boris84,Ioselevich,Zhang2004,Tianran2012,Tianran20121,Tao2006,Tao20061}. In epitaxially connected nanocrystals similar relay co-tunneling happens through necks connecting nearest neighbor nanocrystals~\cite{Brian2012,Han2016,Uwe2016}.

In doped semiconductors the related concept of tunneling with virtual scattering on intermediate impurities was developed in Ref.~\onlinecite{Boris1982,Lifshitz,Transparency,Sasha}. Later it was applied to theories of negative magnetoresistance, Aharonov-Bohm oscillations and mesoscopic transport in the range of ES law~\cite{Vasya1985,NSS1985,NSS19851,BB1990,BB19901}. In QHE the localization length of states between two half-integers is formed by chains of virtually scattered by multiple impurities electron cyclotron orbits~\cite{Fogler}.  

Theory of the variable range hopping thermopower was developed for the Mott law~\cite{Zvyagin} and the ES law~\cite{Chaikin}. The hopping thermopower originates from a small residual asymmetry of the density of localized states~\cite{Zvyagin,Chaikin} or of the localization length~\cite{Yamamoto} near the Fermi level $\mu$ which does not allow electron and hole contributions to completely cancel each other. Experimental confirmation of these theories, however, remains controversial~\cite{Liu, Park, Kaito}. 

\section{Coulomb gap near metal-insulator transition}

The shape of the Coulomb gap of DOS in a doped 3D semiconductor with concentration of donors $N$ close to the critical concentration $N_c$ of the MIT was studied in Ref.~\onlinecite{MarkLee} both theoretically and experimentally. It is known~\cite{SEbook} that on the insulator side of MIT the long wavelength dielectric constant $\kappa$ diverges, so that according to Eq. (\ref{eq:DOS3}) DOS $g(|\epsilon-\mu|)$ parabola becomes very steep close to $N_c$ (see the lower dashed line in Fig. 1). On the other hand, on the metallic side of MIT, $g(\epsilon)$ is large and weakly energy dependent at small $|\epsilon -\mu|$. In order to understand how crossover between these two limits happens it was assumed~\cite{MarkLee} that at MIT dielectric constant is a power law function  of the wave number $q$, namely $\kappa(q) = \kappa (qa)^{1 - \eta}$, where $a$ is the donor Bohr radius. Below we use $\eta = 2$ which is in good agreement with experiments. Then the Coulomb potential of a donor becomes $V(r) = e^{2}a/\kappa r^2$ and using ES arguments of Ref.~\onlinecite{ES1975} we arrive at the critical Coulomb gap DOS, shown in Fig. 1 by the thick line
\begin{figure}[t]
    \centering
    \includegraphics[width=\linewidth]{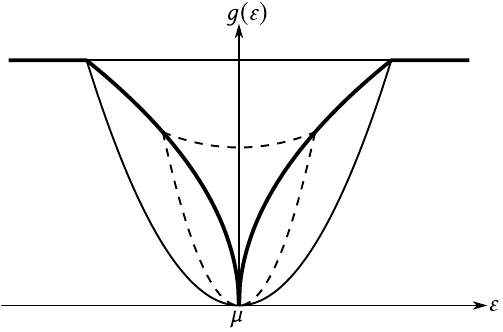}
    \caption{Schematic plot of the DOS $g(\epsilon)$ in the vicinity of the critical concentration of donors $N_c$ of the metal-insulator transition~\cite{MarkLee}.  The DOS $g(\epsilon)$ at $N=N_c$ is shown by the thick line. Thin lines show $g(\epsilon)$ at $N=2N_c$ and $N=N_c/2$. Dashed lines show $g(\epsilon)$ for $N=1.25 N_c$ and $N=0.8 N_c$ at $|\epsilon-\mu | < T_{ES}$ where they deviate from the thick line.}
\end{figure}
\begin{equation} 
g_c(\epsilon) = \frac{\kappa^{3/2}|\epsilon-\mu|^{1/2}}{a^{3/2}e^3}.
\label{eq:criticalgap}
\end{equation} 
At $|1-N/N_c| \ll 1$ the large correlation (localization) length $\xi = a|1-N/N_c|^{-\nu}$ truncates the growth of the dielectric constant $\kappa(q)$ at $\kappa(1/\xi) = \kappa \xi/a$ and defines the energy scale $T_{ES} =  e^2/\xi \kappa(1/\xi) = e^2 a/\kappa \xi^2$. Then at $|\epsilon-\mu| < T_{ES}$ on the insulating side of the MIT we recover Eq.~(\ref{eq:DOS3}) with $\kappa$ replaced by $\kappa \xi/a$, while on the metallic side $g(\epsilon) = g_{c}(T_{ES}) = Const$ (see the upper dash line in Fig. 1). ES law is valid at $T < T_{ES}$ because it uses only the parabolic part of DOS. For typical experimental value $\nu=1$ we get for the ES law $T_{ES} \propto (N_c-N)^2$ in agreement with experiments~\cite{SEbook,McCammon2005}.

 McMillan~\cite{McMillan} arrived essentially at the same scaling ansatz but only for the metallic side of MIT. The McMillan-Shklovskii scaling ansatz was supported by two recent theoretical papers~\cite{Amini,Gornyi}. The Coulomb gap evolution near MIT was studied by direct tunneling into Si crystals~\cite{MarkLee} with impurity concentrations $0.9 N_c < N  < 1.1 N_c$. Results of this experiment are in agreement with McMillan-Shklovskii scaling ansatz.

On the metallic side of MIT transition in Fig. 1 I did not show the electron-electron interaction induced Altshuler-Aronov (AA) DOS dip near the Fermi level ~\cite{AA,AA1}. It was studied theoretically only for a metal with $N >> N_c$ and $k_{F}l >> 1$, where $k_F$ is the Fermi wave number and $l$ is the mean free path. In this case the depth of the AA $|\epsilon-\mu|^{1/2}$ dip is small, $\delta g(\mu) \sim (k_{F}l)^{-3/2}g(\mu) <<  g(\mu)$. However, at $N=2N_c$ one gets $\delta g(\mu) \sim g(\mu)$. As far as I know the behavior of the $\delta g(\mu)$ dip near the MIT was not studied theoretically or numerically. It is likely that a substantial AA dip persists even near the MIT, but only in the shrinking with $N-N_c$ energy range $|\epsilon-\mu| \simeq T_{ES}$. Experimentally studied tunneling DOS on the metallic side of MIT is consistent with this conjecture~\cite{MM,MarkLee}.  

The McMillan-Shklovskii scaling ansatz is based on the absence of the local screening of the Coulomb gap in a given realization of random coordinates of donors and acceptors which was discovered in numerical experiment Ref.~\onlinecite{Baranovskii}, discussed in Ref.~\onlinecite{MarkLee} and later supported by Refs.~\onlinecite{Palassini,Yang}. Indeed, a new electron added to an empty donor D in the bulk of a lightly doped semiconductor induces polarization of either zero or small number of pairs, i. e. transitions of other electrons from occupied to empty donors. These pairs or their chains move the electron charge away from the donor D in a particular direction, so that the potential of the added electron at distance $r$ remains of the order $-e/\kappa r$, but may change its sign. Thus, in a system of localized states the local screening is absent and does not affect the Coulomb gap. To arrive at the exponential Yukawa screening of the Coulomb potential we have to average over many realizations of random coordinates of donors and acceptors surrounding donor D. Such a procedure leads to averaging over many directions of polarization chains and makes the added to D electron's polarization cloud spherically symmetric, but apparently it has nothing to do with the ES law  for a single realization of coordinates of donors and acceptors. On the other hand, when one calculates the thermodynamic DOS describing penetration of the uniform electric field into the large area of orthogonal surface of a macroscopic slab of a lightly doped semiconductor, the averaging of the local polarization over the slab surface makes perfect sense and leads to the finite DOS and the exponential Yukawa screening with the screening radius of the order of average distance between donors~\cite{Baranovskii}.  

So far I have dealt with MIT in 3D semiconductor. It is believed that in a two-dimensional semiconductor there is no MIT with the growing two-dimensional concentration of donors $n_D$. Instead  the localization length $\xi$ of electron states near the Fermi level steeply grows with $n_D$. The Coulomb gap near the Fermi level then can be described assuming that effective dielectric constant $\kappa$ does not change much~\cite{polyakov93} because electric field lines escape to surrounding space. Namely, for the shrinking with $n_D$ energy range $|\epsilon-\mu| < T_{ES}$, where the characteristic distance $e^2/\kappa |\epsilon-\mu|$ is larger than $\xi$, Eq.~(\ref{eq:DOS2}) remains valid, while at $e^2/\kappa |\epsilon-\mu| < \xi$ the Coulomb interaction is so weak that at $|\epsilon-\mu| > T_{ES}$ the DOS sharply returns to its non-Coulomb value. The ES law with $T_{ES}$ decreasing with $\xi$ and $n_D$ then remains valid at all $T < T_{ES}$. These predictions qualitatively agree with the data on tunneling into very thin Be films and on their conductivity~\cite{Butko}. Recently a linear Coulomb gap centered at the Fermi level was observed in tunneling experiments~\cite{Ye,Bruno} in two-dimensional crystals of WTe$_{2}$ and TaTe$_{2}$.

\section{Beyond the Coulomb gap}

\subsection{High frequency conductivity}

In this section I dwell on problems with the Coulomb interaction and disorder which are practically missing in the ES book. As we saw above, the Coulomb gap of the one-electron DOS determines the DC variable range hopping conductivity and the tunneling into a semiconductor. On other hand, the high frequency the phononless hopping conductivity $\sigma(\omega)$ is determined by electron hopping in compact resonance pairs of donors with small arms $r_{\omega}=\xi \ln(E_0/\hbar\omega)$ populated by one electron only. Here $E_0$ is the donor ionization energy. The density of states of these pairs does not have a Coulomb gap and is (with logarithmic accuracy) constant at small excitation energies~\cite{ES1975,SEbook}. So at the first sight the Coulomb interaction is not important here and the low temperature terahertz range phononless conductivity should obey the $\sigma(\omega) \propto \omega^2$ law proposed by Mott~\cite{Mott} and Berezinskii~\cite{Berezinskii}. However, we predicted~\cite{SE1981} that at gigahertz frequencies when the Coulomb energy of two electrons repulsion $e^2/\kappa r_{\omega}$ exceeds $\hbar\omega$ one power of $\hbar \omega$ should be replaced by $e^2/\kappa r_{\omega}$, what leads to $\sigma(\omega) \propto \omega$. This happens because the concentration of pairs having one electron is proportional not to $g(\mu) \hbar\omega$ as assumed by Mott~\cite{Mott} and Berezinskii~\cite{Berezinskii}, but to $g(\mu)e^2/\kappa r_{\omega}$. Multiple experiments~\cite{Hohls,Lewis,Mlee2001,Helgren2002,Helgren2004,Hering} confirmed this prediction.

Above I dealt with the case when $e^2/\kappa r_{\omega}$ is larger than the width of the Coulomb gap $(e^{6}g(\mu)/\kappa^{3})^{1/2}$. In the opposite case, one should take into account the Coulomb gap in the density of states replacing~\cite{SE1981} the density of states $g(\mu)$ by $\kappa/r_{\omega}^{2}e^{2}$. This replacement changes the power of $r_\omega$, but does not modify linear dependence on frequency. 

So far I talked only about zero temperature phononless absorption of photons by compact donor pairs. At finite temperatures and lower frequencies microwave absorption is due to phonon assisted relaxation of electron filling factors of donors of a pair~\cite{Pollak}. The calculation of $\sigma(\omega)$ in Ref.~\onlinecite{Pollak} automatically included interaction between two electrons of a pair of donors, but later in the application to amorphous semiconductors~\cite{Austin} this interaction was ignored. Efros~\cite{EfrosA} took the electron-electron interaction into account for amorphous semiconductors and showed that this interaction weakens the temperature dependence of the relaxation absorption, because one power of $k_B T$ is replaced by $e^2/\kappa r_{\omega}$. Comprehensive review of the theory of both phononless and relaxation high frequency hopping conductivity and dielectric constant can be found in Ref.~\onlinecite{SE1985}. Compact pairs are also responsible for the electron contribution to the specific heat of systems with localized states~\cite{Uzakov}. 

The crossover from the relaxation AC hopping conductivity to the DC hopping conductivity with decreasing frequency happens when the exchange of an electron between donors of a compact pair is replaced by the electron exchange between two growing clusters of donors~\cite{Zvya,Bott}. Eventually, at $\omega \sim 4\pi\sigma(0)/\kappa$ these clusters are incorporated in the percolating infinite cluster of donors responsible for DC transport. 

Far on the other side of the frequency spectrum, at very low frequencies $\omega = 2\pi f << 4\pi\sigma(0)/\kappa$, there is the $1/f$ noise of the hopping conductivity~\cite{McCammon2005}. Trapping and release of electrons by anomalously isolated donors may be responsible for this noise~\cite{onef1,onef3} for not very small $f$, while at smallest $f$ rearrangements of large clusters of interacting electrons are likely to cause the $1/f$ noise~\cite{onef4}. 

\subsection{Back to strong electric fields}

The hopping conductivity in strong electric fields is another big field practically missing in the SE book. In Section III we dealt with very low temperature and very strong electric fields which lead to the linear growth of the effective temperature. However, there are many intermediate cases. At a strong disorder, large concentrations of carriers and low temperatures the current always exponentially grows with the applied electric field~\cite{Shklovskii1976,Shklovskii1979, Levin,SF1,SF2,SF3}. This growth can be interpreted as the effect of reduction of large hopping energy barriers by the applied electric field. However, at small concentrations of carriers, modest disorder and large enough temperatures the current decreases in a strong electric field. This counterintuitive behavior happens because a very strong electric field does not allow electrons to use return parts in optimal tortuous percolation hopping chains responsible for the Ohmic conductivity~\cite{SF4,SF5}. As a result, a strong electric field creates geometrical traps for electrons, such as "parachutes"~\cite{SF4} or "toilets"~\cite{SF5}. Then electrons are forced to spend most of their time in these traps and only rarely leave them. Predicted negative differential conductivity was discovered in lightly doped and weakly compensated Si samples~\cite{SF6}, where it led to new Gunn-effect-like current oscillations. The original theory~\cite{SF4} was confirmed by Monte-Carlo simulations and augmented in Ref.~\onlinecite{NenashevNDC}.

Hopping down an energy ladder of localized states plays an important role in many relaxation phenomena in semiconductors. We studied the energy relaxation via hopping in the nonradiative recombination of electrons and holes~\cite{Rec6,Rec2} and in the hopping photoconductivity~\cite{Rec1,Rec3,Rec4,Rec5}. Spin polarization relaxation via hopping conductivity was discussed in Ref.~\onlinecite{Rec7}. 

\subsection{Wigner Crystal} 

Correlated arrangement of electrons in a lightly doped semiconductor leading to the Coulomb gap is similar to a hierarchy of pinned by disorder Wigner crystals of different spatial scales. Therefore, I returned to Wigner crystals several times. In Refs.~\onlinecite{WC1,WC2,WC4} we studied pinning strength of 2DEG Wigner crystal and the hopping conductivity of its electrons. 

In Refs.~\onlinecite{WC3} we numerically studied the energy of the clean 2D electron Wigner crystal confined in a gated 2D circular quantum dot. We discovered that as number of electrons grows, the Coulomb blockade is accompanied by an additional phenomenon which we term "elastic blockade". It is related to periodic creation or elimination of new Wigner crystal dislocations. The elastic energy of an eliminated dislocation allows two or more electrons to enter the dot together as was earlier observed experimentally~\cite{WC16} (and interpreted by a simple model of three electrons on four localized states~\cite{WC15}). The same physics explains anomalously large ($\sim 15\%$) fluctuations of gate voltage distances between Coulomb blockade peaks or in other words fluctuations of the dot inverse capacitance~\cite{Sivan}. Pinning of such Wigner crystal island by impurities, of course, increases these fluctuations. Their distribution in extreme cases of very strong disorder and its relationship to the Coulomb gap are studied in Ref.~\onlinecite{Pikus}.

Ideas of Wigner crystal were used in theory of supercapacitors~\cite{BTLS}. It was also predicted that a 3D array of epitaxially connected nanocrystals gated by a penetrating inside the array ionic liquid, counterions may abruptly get a finite charge at some critical gate voltage~\cite{Cooper}. At this critical gate voltage the differential capacitance of the array is infinite.

\subsection{Stripes and bubbles in QHE}

While studying spin splitting of Landau levels of 2DEG in a strong magnetic field~\cite{WC5} we were wondering whether electrons of a spin split high Landau level can form a Wigner crystal~\cite{AG}. We realized~\cite{WC6,WC7} that instead of Wigner crystal the peculiar box type of the electron-electron repulsion potential at a partially filled Landau level with quantum number $N \geq 2$ ~\cite{AG}
should lead to the stripe phase at its filling factor $0.4 < \nu < 0.6$ and to bubble phases at $\nu < 0.4$ and $\nu > 0.6$. Predicted in~\cite{WC6,WC7}  phases were discovered in Refs.~\onlinecite{Jim,Du}. See a comprehensive review of the first 6 years of stripes and bubbles~\cite{WC8}. 

Following ideas of Ref.~\onlinecite{WC9} we recently calculated the ratio of large and small resistivities in a stripe phase and found that it should grow with the magnetic field $B$ as $B^{4}$. This result is in agreement with the experimental data for cleanest GaAs samples where the resistance anisotropy reaches 3000~\cite{WC10}. We also studied what the scattering by a short range disorder does to a stripe phase and its resistivity tensor. We discovered "hidden stripe phases" where the stripe order survives a growing short range disorder, but the resistivity becomes isotropic~\cite{WC11,WC12}.

Multiple bubble phases with different numbers of electrons per bubble were recently observed in GaAs~\cite{Zudov,Csathy} and graphene~\cite{Dean,Andrea}. They are direct relatives of Wigner crystals observed at $\nu < 1/5$ and $\nu > 4/5$. 

\subsection{Negative compressibility}

Theory~\cite{Bello,Chenskii} predicted that the compressibility of a low-density 2DEG in the Wigner crystal and Wigner-crystal-like liquid states is negative. When a 2DEG with a very small concentration $n$ is on one side of a capacitor with opposite metallic electrode at the distance $d$ this negative compressibility leads to small negative correction for $d$, namely $d$ is replaced by $d^*= d-0.12n^{-1/2}$.

The negative compressibility was discovered in Refs.~\onlinecite{Krav,Eisen} and studied in Refs.~\onlinecite{Shapira,Dultz,Ilani,Savchenko,Tituc}. Many authors working with relatively small $r_s = (\pi n)^{-1/2}a_B^{-1} < 5$, where $a_B$ is a semiconductor Bohr radius, found a satisfactory agreement of their data with the result of the Hartree-Fock approximation and claimed that the negative compressibility is the exchange driven phenomenon. Indeed, exchange makes compressibility negative~\cite{Ceperley} starting at $r_s > 2.2$ and leads to $d^* = d-0.064n^{-1/2}$. Thus, the exchange negative compressibility correction is roughly twice smaller than in the Wigner crystal limit. Actually both the Pauli principle driven exchange between same spin electrons and the direct Coulomb repulsion between all electrons keep electrons apart reducing their energy. At $r_s > 5$ the role of the exchange energy decreases, while the role of the correlation energy increases, so that at $r_s > 10 $ the energy of the liquid state is practically equal to that of the Wigner crystal, where the exchange contribution is exponentially small. 

Thinking about the Wigner crystal and Wigner-crystal-like liquid at $r_s > 10$ we studied~\cite{Skinner2010} what happens at a very small $d$ when $nd^2 \ll 1$. We found that in this case $d^*= 2.7d (nd^2)^{1/2} \ll d$, so that capacitance becomes much larger than geometrical. This limit has not been achieved yet, but 40$\%$ and 60$\%$ growth of the capacitance above geometrical was observed in Refs.~\onlinecite{Ashoori,Mak}. (References~\onlinecite{Bello,Skinner2010} dealt with an insulator with isotropic dielectric constant, but one can show that their results for $d^*$ work also for a plane capacitor with dielectric constant in its plane is much larger than in the orthogonal to it direction, where the geometrical capacitance is determined by the latter one.)

Strong magnetic field enhances the negative compressibility near integer filling factors where the concentration of electrons or holes at the topmost Landau level is very small~\cite{Krav,Eisen}. This effect was studied theoretically in GaAs~\cite{Singapore,Skinner2013} and graphene~\cite{Manchester}. Capacitance  $20\%$ larger than geometrical was observed in graphene in strong magnetic field~\cite{Manchester}.                                                 
On the other hand, a strong disorder destroys Wigner crystal correlations at small electron densities and makes the compressibility positive~\cite{Eisen,Skinner2013}.

\subsection{Charge inversion of DNA and gene delivery}

We also used the concept of a classical Wigner crystal beyond solid state physics, namely in the chemistry and biology of water solutions with multivalent positive ions, for example, short positive polyelectrolytes. We showed that the Wigner-crystal-like correlated liquid of multivalent counterions adsorbed at the surface of a negative macroion such as the double-helix DNA inverts the DNA net charge making DNA positive~\cite{BIS,CI1,Loth}. This charge inversion is driven by attraction of extra counterions to their correlation holes (images) in the surface of DNA already neutralized by absorbed Wigner-crystal-like liquid of counterions. The net positive charge of DNA grows until it's repulsion compensates the counterion-hole attraction. Charge inversion can be also interpreted as a result of the classical fractionalization of the polyelectrolytes driven by elimination of its self energy~\cite{CI2}. We deal with a multivalent strongly charged counterions because repulsion energy of monovalent ions on DNA surface is much weaker and they can not form  Wigner-crystal-like correlated liquid at room temperature.

Membranes of biological cells are typically negative and repel approaching strongly negative naked DNA molecules, while a charge inverted positive DNA is allowed to adsorb and enter the cell. Therefore, DNA charge inversion by positive polyelectrolytes is used for the delivery of normal genes to the cells with abnormal genes producing defective proteins~\cite{CI3,CI4}. This paradigm of the modern medicine is called the gene therapy. Charge inversion of a mRNA molecule by positive lipids is used in modern COVID-19 Pfizer and Moderna mRNA vaccines~\cite{Langer}. 

Nature uses Coulomb interaction of negatively charged DNA and RNA to positive tails of some proteins for construction of chromatin and viruses. We studied simple models of kinetics of Coulomb interaction guided self-assembly of vegetable viruses and estimated assembly time for different contents of nucleic acids and proteins~\cite{Tao}.

In another our biology related theory we solved the puzzle of the deconfinement of pairs of positive and negative salt ions in narrow biological ion channels 
which leads to practically zero activation energy of the channel ion conductivity at ambient conditions~\cite{IonC1,IonC2}. 
This deconfinement is similar to the one of quarks in a hot quark-gluon plasma produced in high energy heavy ion collisions~\cite{Susskind}.

\section{SE book at 40. Comments to its chapters} 

Chapter 1. Electron binding energy of a hydrogen-like donor in a bilayer graphene with a gate-tunable gap was studied in Ref.~\onlinecite{Edik}. 

Chapter 2. Substantial progress in identifying an MIT point and it's critical indexes is achieved by analyzing the distribution function $P(s)$ of the nearest level spacing $s$. It was conjectured and confirmed numerically~\cite{Shore93} that on the top of two known universal functions $P(s)$, the Wigner-Dyson and the Poisson ones, exactly at the MIT there is the third universal function $P(s) = P_{c}(s)$. It is an interesting hybrid of the first two. Namely, it was found to behave as the Wigner-Dyson one at $s < 1$ and as the Poisson $P(s) \propto \exp(-As)$ at $s>>1$. Here $A > 1$ due to the
level repulsion. Later study of much larger 3D lattices~\cite{Isa} confirmed the $s>> 1$ asymptotic of $P_{c}(s)$ with $A=1.9 \pm 0.1$.  

We showed that for the Anderson transition one can characterize the function $P(s, W, L)$ for each disorder strength $W$ and the system size $L$ by the area $\gamma(W, L)$ of its $s > 2 $ tail. Then doing the finite size analysis for the function $\gamma(W, L)$ we found the transition point $W_c$ and the correlation length exponent~\cite{Shore93}. Thus, to find an MIT point one does not need to study the participation ratio of wave functions, but can study only the energy spectrum. The finite size analysis of $\gamma(W, L)$ was used to identify MIT transitions for many years, until a new more convenient method of characterizing $P(s, W, L)$ called $r$-method was invented~\cite{Oganesyan}. One can also use the spectral rigidity~\cite{Dyson,Altshuler,FoglerD} to find an MIT~\cite{Kot88}. 

Recently we extended the finite size analysis to the non-Hermitian analog of Anderson Hamiltonian where complex diagonal elements are distributed randomly in a square with the side $W$ in the complex plane~\cite{Makris}. Using generalization of the $r$-method for analysis of complex energies~\cite{Prosen} we discovered the non-Hermitian Anderson transition in three-dimensions and showed that in the non-Hermitian case as in Hermitian one, there is no Anderson transition in two-dimensions~\cite{Yi20,Yi2020}.

Classical localization of 2D electron gas in a long range random potential with increasing magnetic fields was studied in Ref.~\onlinecite{Perel}. It was shown that it results in the positive magnetoresistance~\cite{Mirlin}. 

Chapter 3. Calculation of the activation energy of the nearest neighbor hopping conductivity of a 3D weakly compensated semiconductor~\cite{SEbook,Tkach} was generalized for the 2D case~\cite{Nelson15}. The goal was to describe the experimental data for the surface hopping conductivity of Si induced by ion liquid gating with a positive metallic gate which pushes positive ions of the ion liquid to Si surface. Inside of the Si crystal these ions play the role of surface donors which due to compensation by bulk acceptors provide surface hopping conductivity. 

Chapter 4. Hall effect for the hopping conductivity was theoretically studied and compared with other theoretical results in Ref.~\onlinecite{Galperin}.

Chapter 5. Percolation theory was reviewed with emphasis on finite cluster scaling and fractal dimensions in the excellent book~\onlinecite{Stauffer}. A new application of percolation theory was proposed~\cite{Orth,Orth1} to describe bulk magnetization induced by ion liquid gating of an oxide film. 

Chapter 6. Second section of Chapter 6 deals with the anisotropy of the nearest neighbor hopping conductivity originating from the anisotropy of the electron spectrum and donor wave functions. Here we would like to discuss the Coulomb gap and ES law in anisotropic systems. Let us start from an anisotropic crystal with randomly distributed in space donors and with different diagonal dielectric constants $\kappa_x$, $\kappa_y$, $\kappa_z$ and localization lengths $\xi_x$, $\xi_y$, $\xi_z$. 
To see how this changes the Coulomb gap one can introduce new coordinates $x'= x\kappa_x^{-1/2}$, $y'= y\kappa_y^{-1/2}$, $z'= z\kappa_z^{-1/2}$ in which the Coulomb potential of a charge $e$ is isotropic~\cite{LL}: $\phi(r') = e/(\kappa_{x} \kappa_{y} \kappa_{z})^{1/2} r'$. Then in the $(x',y',z')$-system,
the Coulomb gap is given by  Eq. (\ref{eq:DOS3}) with $\kappa^{3}$ replaced by $(\kappa_x\kappa_y\kappa_z)^{3/2}$. Returning back to the original $(x, y, z)$-system one arrives at the Coulomb gap DOS Eq. (\ref{eq:DOS3}) with $\kappa^{3}$ replaced by $\kappa_x\kappa_y\kappa_z$.

Let us now discuss changes in the ES law starting from the $(x',y',z')$-system where the dielectric constant is isotropic and equal $(\kappa_x\kappa_y\kappa_z)^{1/2}$, but the localization length remains generally speaking anisotropic: $\xi_{x}' = \xi_{x} \kappa_x^{-1/2}$, $\xi_{y}' = \xi_{y} \kappa_y^{-1/2}$, $\xi_{z}' = \xi_{z} \kappa_z^{-1/2}$. According to Chapter 6 this means that in the $(x',y',z')$-system we arrive at the ES law with the temperature $T_{ES}$ where the product $\kappa \xi$ is replaced by $(\kappa_x\kappa_y\kappa_z)^{1/2} (\xi_x'\xi_y'\xi_z')^{1/3} = (\kappa_x\kappa_y\kappa_z\xi_x\xi_y\xi_z)^{1/3}$. Going back to the laboratory $(x,y,z)$-system does not change the exponential term of the ES law determined by the percolation threshold (see Chapter 6), while the anisotropy of the final prefactor is determined only by the anisotropy of the localization length, i. e. $\sigma_{xx}/\xi_x^2 = \sigma_{yy}/\xi_y^2 = \sigma_{zz}/\xi_z^2$. 

Situation is different when donors are randomly distributed only in a periodic stack of parallel $xy$-planes separated by a large distance $d$ along the $z$-axis. Then the above theory is correct and the conductivity anisotropy is non-exponential only at low enough temperatures when the typical electron hop length $r_h \gg d$ and the ES law works in all directions. At higher temperatures the larger in-plane conductivity is due to ES variable range hopping with $r_h \ll d$, while the much smaller $z$-axis conductivity is due to the hopping with $r_h \simeq d$. It is likely that the resulting exponentially large anisotropy of the hopping conductivity was recently discovered in Ref.~\onlinecite{Moser7}. 
 
Chapter 12. SE book section (12.3) deals with  filtering of a random potential from its small scales via the wave function averaging. Recently a new way to do the filtering was suggested~\cite{Sveta} and later compared with the conventional one~\cite{Serezha1}.

Chapter 13. The theory of nonlinear screening of a long-range potential developed in Chapters 3 and 13 was extended to nonlinear screening of the random potential of remote charged impurities by the 2DEG in QHE devices~\cite{DOS,WC5,Efros1,Efros2,Efros3,CSG,CMS,Fogler94,Sammon,Sammon1}.
The concepts of the random long range potential and percolative metal-insulator transition in strongly or completely compensated semiconductor found applications to topological insulators~\cite{Why,Ando,Surface,TopolFilm,TopolWire}, to graphene~\cite{Galitski}, to two-dimensional narrow gap semiconductors subjected to a strong 3D Coulomb disorder~\cite{Narrowgap,Smallgap}, to Ge and ScN films~\cite{Mitin,Biswas}, to MIT in doped SrTiO$_3$ crystals and films~\cite{Yilikal} and to the superconductor-insulator transition in high $T_c$ superconductors~\cite{Compsuper,Segawa}.

In $\text{SrTiO}_3$ due to its closeness to the ferroelectric transition nonlinear dependency of polarization on electric field was added to the standard Thomas-Fermi nonlinear dependence of electron concentration on the electrostatic potential to analytically describe low temperature thermodynamic and kinetic properties of 2DEG in LAO-STO junctions in zero magnetic field~\cite{Kostya,HanKostya,HanMike} and in a very strong magnetic field~\cite{MikeHan}.
The same theory was applied to collapse of electron gas on a positive cluster of oxygen vacancies in $\text{SrTiO}_3$~\cite{KostyaHan}.

This paper is devoted to the memory of Sergey Gredeskul and his wife Victoria murdered by Hamas in their home. Sergey was fascinated by the phenomenon of Anderson localization in disorder systems. We both followed seminal works of I. M. Lifshits. Sergey's book written with I. M. Lifshits and L. A. Pastur~\cite{Gredeskul} and SE book~\cite{SEbook} augment each other. 

\begin{acknowledgements}
I am grateful to all my coworkers and particularly to A. L. Efros for enjoyable collaborations. 


\end{acknowledgements}

\end{document}